# Effect of image charges on differential trajectories measurements

A. Shemyakin, Fermilab, Batavia, IL 60510, USA
*shemyakin@fnal.gov*

*Abstract*
Effect of image charges on reconstruction of focusing elements calibration with differential trajectory method is estimated and found negligible for measurements at the PIP2IT MEBT.

## 1. Introduction

A standard way of measuring the linear optics of an accelerator or a beam line is the differential trajectory method:
- The nominal beam trajectory is recorded with Beam Position Monitors (BPMs)
- The beam is deflected with a dipole corrector so that the deflected trajectory goes through the linear region of focusing elements
- The difference between the deflected and nominal trajectories is compared with predictions of the optical model. Found discrepancies usually indicate errors in connections, positions, or calibration of the magnets.

Interpretation of the results of the differential trajectory measurements normally neglects interaction of the beam with image charges at the vacuum pipe. This note estimates this effect in simple models. Numerical estimations are made for the case of measurements at the Medium Energy Beam Transport line (MEBT) of the PIP2IT test accelerator [1].

Since strength of interaction with image charges drops quickly with particle energy, all consideration is for the non-relativistic case. Image currents are neglected, and only electrostatic forces are taken into account.

## 2. The force

The first step is to estimate the force acting on beam particles from the charges at the vacuum pipe walls (image charges) when the beam is slightly off-axis. The pipe is considered to be an infinite conductive cylinder of radius $R$, with the beam transverse size $\ll R$ ("pencil beam").

- *DC beam*

The pencil beam of particles with charge $e$ and mass $m$ moving with the velocity $v$ at the offset $x$ from the axis (Fig.1) is approximated by a thin charged wire with the constant linear charge density $\rho = I_b/v$, where $I_b$ is the beam current.

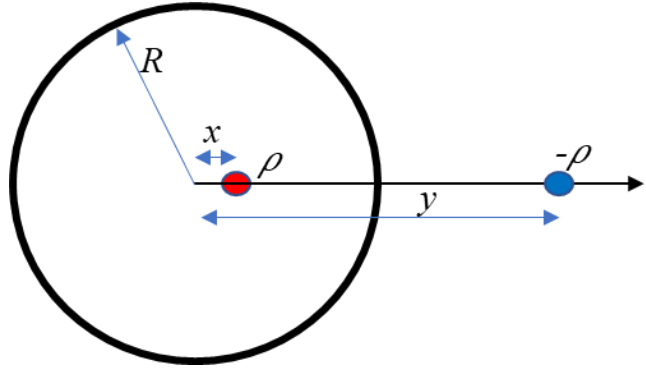

Figure 1. Geometry of estimation for the DC beam case.

Inside the cylinder, electric fields from the image charges are equivalent to the fields from the line charge $-\rho$ placed at the distance $y = R^2/x$ from the axis. Correspondingly, the force acting on a beam particle is

$$F_{DC}(x) = \frac{e \cdot I_b}{v} \frac{2}{y-x} \approx \frac{e \cdot I_b}{v} \cdot \frac{2x}{R^2} \qquad (1)$$

- *Point-like bunches*

Let's consider the case of a beam consisting of bunches which sizes are much smaller than the pipe radius (Fig.2).

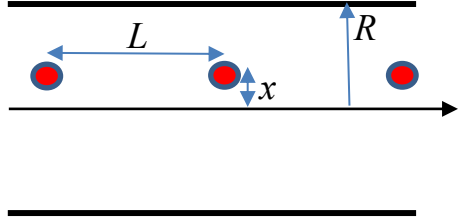

Figure 2. Geometry of estimate for the case of point-like bunches.

When the separation between bunches is large $L = v/f \gg R$, where $f$ is the bunch frequency, force from the image charges of neighboring bunches can be neglected. The force on a beam particle from image charges induced by a point-like bunch with the bunch charge $Q_b$ follows from Eq. (23) in Ref. [2]

$$F_p(x) = \frac{2eQ_b}{\pi} \sum_{m=-\infty}^{\infty} \int_0^\infty k \cdot I_m(kx) I_m(kx)' \frac{K_m(kR)}{I_m(kR)} dk, \qquad (2)$$

where $I_m$ and $K_m$ are modified Bessel functions of the first and second kind of $m$-th order. For the case of $x \ll R$, the sum is quickly converging, and its numerical evaluation gives

$$F_p(x) \approx \frac{eQ_b}{R^3} x \qquad (3)$$

For comparison with Eq. (1), the bunch charge in Eq.(3) can be expressed through the average beam current as $Q_b = I_b/f$

$$F_p(x) \approx \frac{eI_b}{R^3 f} x \approx F_{DC}(x) \cdot \frac{L}{2R} \quad (4)$$

- *Discussion of a general case*

For a case of a large separation but finite bunch length, the force can estimated using Eq.(19) from Ref. [2] after removing the component related to self-field. For a Gaussian distribution with rms bunch length $\sigma$, the force on the central particle is

$$F_p(x) = \frac{2eQ_b}{\pi} \int_{-\infty}^{\infty} \frac{e^{-\frac{z^2}{2\sigma^2}}}{\sqrt{2\pi}\sigma} dz \sum_{m=-\infty}^{\infty} \int_0^{\infty} k \cdot \cos(kz) I_m(kx) I_m(kx)' \frac{K_m(kR)}{I_m(kR)} dk . \quad (5)$$

Changing order of integration gives

$$F_p(x) = \frac{2eQ_b}{\pi} \sum_{m=-\infty}^{\infty} \int_0^{\infty} k \cdot e^{-\frac{k^2\sigma^2}{2}} \cdot I_m(kx) I_m(kx)' \frac{K_m(kR)}{I_m(kR)} dk . \quad (6)$$

Numerical calculation of Eq. (6) is shown in Fig.3 as a function of $\sigma/R$ with the force normalized by the value in Eq.(3). Correction is <10% at $\sigma/R < 0.35$.

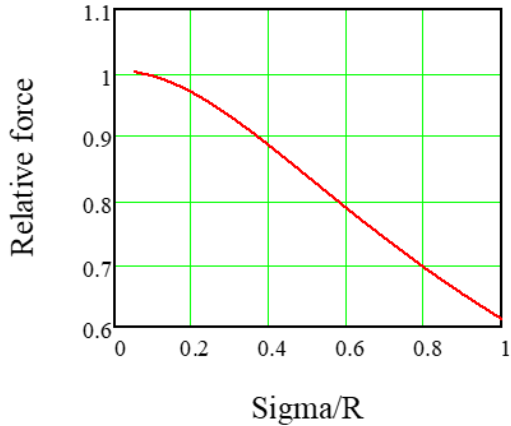

Figure 3. Dependence of the force from image charges imposed on the central particle on the relative rms width of the bunch.

When $\sigma \sim R$, the force from image charges on central and tail particles differs significantly. Since BPMs report values averaged over the bunch, estimates become more complicated, and this case is not considered here. Note that if the force is large, this effect may result in distortion of the bunch phase portrait and an increase of the longitudinal emittance.

For an arbitrary separation between bunches, in the paraxial approximation one should expect the same linear dependence of the image charge force on the offset

$$F(x) \approx \kappa \cdot x \quad (7)$$

with the coefficient $\kappa$ being between predictions by Eq.(1) and Eq.(3). Corresponding expression for an infinite string of bunches separated by distance $L$ comes from summation of Eq.(5) over image charges of all bunches

$$F(x) = \frac{2eQ_b}{\pi} \sum_{n=-\infty}^{\infty} \sum_{m=-\infty}^{\infty} \int_0^\infty k \cdot \cos(knL) \cdot e^{-\frac{k^2\sigma^2}{2}} \cdot I_m(kx)I_m(kx)' \frac{K_m(kR)}{I_m(kR)} dk \ . \quad (8)$$

Numerical calculation of Eq.(8) with 5 Gaussian bunches ($n= -2\ldots2$) is not monotonical at large *L/R*, likely because of a numerical noise, but otherwise shows the expected behavior (Fig. 4). When the separation is less than the pipe diameter, the force is close to Eq.(1) (DC beam), while for a larger separation it is described by Eq.(3).

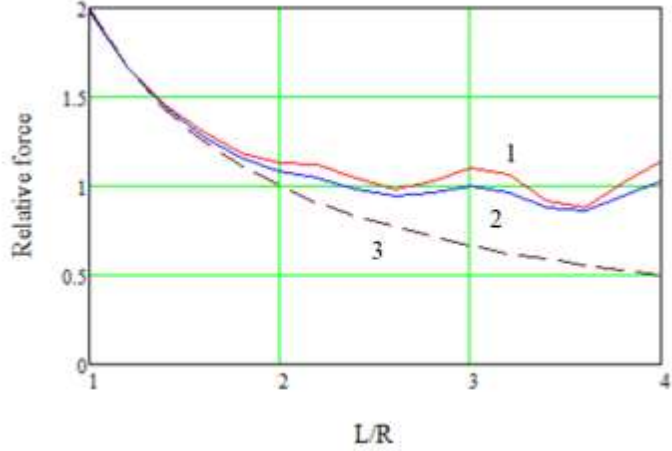

Figure 4. Dependence of the force from image charges imposed on the central particle on the relative distance between bunches. The force is normalized by the value for a single point-like bunch in Eq.(3). Curves 1 and 2 are numerical calculation for 5 Gaussian bunches with the ratio of $\sigma/R=$ 0.01 (1) and 0.3 (2). Curve 3 is the coefficient *2·R/L* from Eq. (4) that in this normalization gives the behavior for the DC case of Eq.(1).

Therefore, with accuracy of ~10% the coefficient in Eq.(7) can be approximated as

$$\kappa \approx \begin{cases} \frac{e \cdot I_b}{v} \cdot \frac{2}{R^2}, & L \leq 2R \\ \frac{e \cdot I_b}{v} \cdot \frac{L}{R^3}, & L > 2R \end{cases}. \quad (9)$$

## 3. Effect of image charges on differential trajectory measurements

- *Constant focusing*

Let's consider a pencil beam propagating inside a metal cylinder through a constant-focusing channel with Twiss beta-function $\beta_0$ at zero beam current. Equation of motion of the beam centroid is

$$x'' + \left(\frac{1}{\beta_0^2} - \frac{1}{\lambda^2}\right)x = 0 \quad (10)$$

where $\lambda^2 = \frac{mv^2}{\kappa}$ and $\kappa$ is the coefficient from Eq.(9). For $\lambda > \beta_0$, a paraxial trajectory is still sinusoidal with increased wavelength so that the new beta-function is

$$\beta_1 = \frac{\beta_0}{\sqrt{1-\left(\frac{\beta_0}{\lambda}\right)^2}}. \qquad (11)$$

For $\lambda^2 \gg \beta_0^2$, fitting the trajectory would predict focusing weaker by

$$\frac{\beta_1^2}{\beta_0^2} - 1 = \left(\frac{\beta_0}{\lambda}\right)^2. \qquad (12)$$

For a real beam line, Eq.(10) should give a reasonable estimate if a typical value for the beta-function is used.

- *Thin lenses (FOFO)*

Another simple model is a beam line with thin focusing lenses with focal length $F$ separated by drifts $S$. The matrix of the period is

$$M = \begin{pmatrix} \cosh\left(\frac{S}{\lambda}\right) & \lambda \cdot \sinh\left(\frac{S}{\lambda}\right) \\ \frac{\sinh\left(\frac{S}{\lambda}\right)}{\lambda} & \cosh\left(\frac{S}{\lambda}\right) \end{pmatrix} \cdot \begin{pmatrix} 1 & 0 \\ -\frac{1}{F} & 1 \end{pmatrix} = \begin{pmatrix} \cosh\left(\frac{S}{\lambda}\right) - \frac{\lambda}{F}\cdot \sinh\left(\frac{S}{\lambda}\right) & \lambda \cdot \sinh\left(\frac{S}{\lambda}\right) \\ \frac{\sinh\left(\frac{S}{\lambda}\right)}{\lambda} - \frac{\cosh\left(\frac{S}{\lambda}\right)}{F} & \cosh\left(\frac{S}{\lambda}\right) \end{pmatrix}. \qquad (13)$$

The phase advance over the period is

$$\cos(\mu) = \frac{m_{11}+m_{22}}{2} = \cosh\left(\frac{S}{\lambda}\right) - \frac{\lambda}{2\cdot F}\cdot \sinh\left(\frac{S}{\lambda}\right) \approx 1 - \frac{S}{2\cdot F} + \frac{S^2}{2\cdot \lambda^2}\left(1 - \frac{S}{6\cdot F}\right). \qquad (14)$$

To fit to the measured phase advance, the fitting procedure of the differential trajectory measurements needs to assume the focusing strength of the lenses being decreased by

$$\frac{\Delta F}{F} \approx \frac{F\cdot S}{\lambda^2}\left(1 - \frac{S}{6\cdot F}\right). \qquad (15)$$

- *FODO*

Derivation for the case of thin lenses with alternating focusing (+F/-F) and distance between lenses of S (i.e. FODO period of 2S) is similar to the case above. The analog of Eq.(15) is

$$\frac{\Delta F}{F} \approx 2\left(\frac{F}{\lambda}\right)^2\left[1 - \frac{1}{12}\left(\frac{S}{F}\right)^2\right] \qquad (16)$$

## 4. Estimates for PIP2IT MEBT

In PIP2IT MEBT, the beam of 2.1 MeV H$^-$ ions is focused transversely by quadrupole triplets with distance between triplet centers of $S=1.175$ m. Typical transverse rms beam size is 2 mm, while the vacuum pipe radius is $R=15$ mm, so the approximation of the pencil beam in Section 2 is valid.

The average beam current is 5 mA. The beam is bunched at 162.5 MHz, which corresponds to a large ratio of the separation between bunches to the tube radius, $L/2R = 123mm/30mm = 4.1$, so Eq.(3) is applicable. With the typical rms bunch length of 0.2 ns or $\sigma$=4 mm $<<R$, the bunch is close to a point-like. In this case, the expression for the image charge-associated length can be written as

$$\lambda = \lambda_{DC}\sqrt{\frac{2R}{L}} = R\left(\frac{mv^3}{2eI_b}\right)^{1/2}\sqrt{\frac{2R}{L}} = R\sqrt{\frac{R}{L}\beta_r^3\frac{I_0}{I_b}} \qquad (17)$$

where $\lambda_{DC}$ is the image charge-associated length for the DC beam, $\beta_r = v/c$ is the relativistic factor, and $I_0 \equiv \frac{mc^3}{e}$ is the characteristic current (3.1·10$^7$ A for protons or H$^-$).

For the PIP2IT MEBT, $\lambda$ =5.1 m, and Eq.(15) gives $\frac{\Delta F}{F} = 0.9\%$, which is below of the present accuracy of the measurements. Note that Eq.(12) gives a reasonable estimate of the effect (~1%) when the average β-function at the MEBT (~0.8 m) is used.

## 5. Summary

Effect of image charges on reconstruction of focusing elements calibration with the differential trajectory method is estimated. Expression for the characteristic length associated with image charges is derived. For the PIP2IT MEBT measurements, the effect is found to be negligible.

## 6. Acknowledgement